# New Particle and Energy "Timeless Particle" and "Timeless Energy" that Existed Before the Beginning of Time

**Ali Riza AKCAY**
Barbaros Mah. Veysipasa Sok.
100. Yil Sitesi H-Blok No: 17/5
81150 Uskudar, Istanbul TURKEY
akcayar@e-kolay.net
(May 17th, 2004)

**Abstract**
This paper gives preliminary information regarding the new particles and energies titled "Timeless Particle" and "Timeless Energy" that existed before the beginning of time.

## 1. INTRODUCTION

The "Conversion Theory" is a new theory that can predict and describe everything on the beginning, evolution and end of time, particles, matters and the Universe. This paper describes the new particles and energy titled as "timeless particle" and "timeless energy" that existed before the beginning of time. According to the "Conversion Theory", there were "timeless particles" and "timeless energy" in the Universe before the beginning of time, and there are and there will be after the end of time the "timeless particles" and "timeless energy" in the Universe. Then, a part of these particles and energy has been converted to the time-dependent particles and time-dependent energy. The time has began just after this conversion. It is quite clear that the terms of antiparticles and negative energy are not clear and have not been described well. In addition, there are a lot of contradictions on the definitions of these terms, and these definitions cannot predict and describe everything in the Universe. The "Conversion Theory" ("CT") describes that there are no antiparticles and negative energy in the Universe. But, there are timeless particles and timeless energy in the Universe. This means that "CT" describes time-dependent particles in place of particles, timeless particles in place of antiparticles, time-dependent energy in place of positive energy and timeless energy in place of negative energy. Please note that the terminologies of "timeless particles" and "timeless energy" are completely new and are used first time in the "CT". Any time-dependent particle and energy cannot observe and detect any timeless particle and energy but, any timeless particle and energy can observe and detect any time-dependent particle and energy. According to the "CT" timeless particles can be converted to time-dependent particles and vice versa, and timeless energies can be converted to time-dependent energies and vice versa.

## 2. CONCLUSION

The preliminary information regarding new particles and energies has been given in this paper.